%% file: main.tex
\pdfoutput=1
\documentclass[manuscript,screen]{acmart}

\newcommand{\italquote}[1]{\begin{quote}``\textit{#1}''\end{quote}}

\AtBeginDocument{%
  \providecommand\BibTeX{{%
    \normalfont B\kern-0.5em{\scshape i\kern-0.25em b}\kern-0.8em\TeX}}}
\newcommand{\revision}[1]{\textcolor{black}{#1}}

\newenvironment{revisionenv}
  {\par\color{black}}
  {\par}

\newcommand{\revisionsecond}[1]{\textcolor{black}{#1}}

\newenvironment{revisionenvsecond}
  {\par\color{black}}
  {\par}

\setcopyright{acmcopyright}
\copyrightyear{2018}
\acmYear{2018}
\acmDOI{XXXXXXX.XXXXXXX}

\acmConference[Conference acronym 'XX]{Make sure to enter the correct
  conference title from your rights confirmation emai}{June 03--05,
  2018}{Woodstock, NY}
\acmPrice{15.00}
\acmISBN{978-1-4503-XXXX-X/18/06}

\begin{document}

\title[User-Centered Design of an Application to Promote Gratitude]{``Actually I Can Count My Blessings'': User-Centered Design of an Application to Promote Gratitude \revisionsecond{Among Young Adults}}

\author{Ananya Bhattacharjee}
\affiliation{%
  \institution{Computer Science, University of Toronto}
  \city{Toronto}
  \state{Ontario}
  \country{Canada}
}
\email{ananya@cs.toronto.edu}

\author{Zichen Gong}
\affiliation{%
  \institution{Intelligent Adaptive Interventions Lab, University of Toronto}
  \city{Toronto}
  \state{Ontario}
  \country{Canada}
}

\author{Bingcheng Wang}
\email{bingcheng.wang@mail.utoronto.ca}
\affiliation{%
  \institution{Intelligent Adaptive Interventions Lab, University of Toronto}
  \city{Toronto}
  \state{Ontario}
  \country{Canada}
}

\author{Timothy James Luckcock}
\affiliation{%
  \institution{Queen Mary University of London}
  \city{London}
  \country{United Kingdom}
}

\author{Emma Watson}
\affiliation{%
  \institution{University College London}
  \city{London}
  \country{United Kingdom}
}

\author{Elena Allica Abellan}
\affiliation{%
  \institution{University College London}
  \city{London}
  \country{United Kingdom}
}

\author{Leslie Gutman}
\affiliation{%
  \institution{Applied Development and Health Psychology, University College London}
  \city{London}
  \country{United Kingdom}
}

\author{Anne Hsu}
\affiliation{%
  \institution{Queen Mary University of London}
  \city{London}
  \country{United Kingdom}
}

\author{Joseph Jay Williams}
\affiliation{%
  \institution{Computer Science, University of Toronto}
  \city{Toronto}
  \state{Ontario}
  \country{Canada}
}
\email{williams@cs.toronto.edu}


\renewcommand{\shortauthors}{Bhattacharjee et al.}

\begin{abstract}
  \input{Sections/abstract}
\end{abstract}

\ccsdesc[500]{Human-centered computing~Empirical studies in HCI}

\keywords{gratitude, psychological wellbeing, user-centered design, mental health, mobile phone, young adults}

\maketitle

\input{Sections/intro}
\input{Sections/related_work}
\input{Sections/overview}
\input{Sections/formative_procedure}

\input{Sections/formative_results.tex}
\input{Sections/deployment_procedure}
\input{Sections/deployment_results}

\input{Sections/discussion}
\input{Sections/conclusion}
\bibliographystyle{ACM-Reference-Format}
\bibliography{sample-base}
\end{document}

%% file: Sections/abstract.tex

Regular practice of gratitude has the potential to enhance psychological wellbeing and foster stronger social connections among young adults. However, there is a lack of research investigating user needs and expectations regarding gratitude-promoting applications. To address this gap, we employed a user-centered design approach to develop a mobile application that facilitates gratitude practice. Our formative study involved 20 participants who utilized an existing application, providing insights into their preferences for organizing expressions of gratitude and the significance of prompts for reflection and mood labeling after working hours. Building on these findings, we conducted a deployment study with 26 participants using our custom-designed application, which confirmed the positive impact of structured options to guide gratitude practice and highlighted the advantages of passive engagement with the application during busy periods. Our study contributes to the field by identifying key design considerations for promoting gratitude among young adults.

%% file: Sections/intro.tex
\section{Introduction}

In today's fast-paced and technology-driven world, young adults are facing increasing levels of stress and anxiety. The pressures of academic and career success, social media, and personal relationships can often lead to feelings of burnout, loneliness, and low self-esteem \cite{bhattacharjee2022kind, o2017design, kornfield2022meeting}. Additionally, many young adults may lack the time or resources to engage in activities that promote their mental and physical wellbeing \cite{bhattacharjee2022kind, bhattacharjee4057942exploring}. The inherent transitional nature of this life stage, encompassing significant changes in education, career, and personal relationships, renders young adults particularly vulnerable to mental health challenges. \revision{Statistically, young adults in the 18--25 age group have the highest prevalence of mental illness, with around 33.7\% experiencing such conditions, significantly higher than older age groups \cite{nimh2020mentalhealth}. Furthermore, there has been a marked increase in symptoms consistent with major depression among this demographic 
\cite{apa2017mentalhealth}. During challenging times like the COVID-19 pandemic, nearly half of these young adults struggled with mental health issues, with many unable to access necessary care \cite{ucsf2021mentalhealth}.} Consequently, there is a growing need for accessible and effective interventions that can help young adults manage stress, improve their general wellbeing, and foster positive social connections. In response to this need, practicing gratitude has emerged as a potential solution to the wellbeing challenges faced by young adults \cite{fritz2019gratitude, lyubomirsky2008happiness}. Research studies have consistently shown that regularly practicing gratitude can lead to a range of positive outcomes, including improved mental and physical health, stronger relationships, and greater life satisfaction \cite{breen2010gratitude, brown2003measuring, roberts2004blessings, watkins2004gratitude, lampinen2013indebtedness, lyubomirsky2008happiness}.

Despite the well-documented benefits of gratitude practice, incorporating it into their daily lives can be challenging for young adults \cite{bhattacharjee2022design, lyubomirsky2008happiness}. These challenges can include being unfamiliar with the potential value of gratitude, a lack of guidance on effective gratitude practice, and time constraints \cite{bhattacharjee2022connecting, kornfield2020energy, bhattacharjee2023investigating, alqahtani2022usability, lyubomirsky2008happiness}. \revision{However, the existing literature offers limited insights into how specific features of gratitude-promoting applications can cater to the varied challenges and contextual needs of users. Previous tools have been typically designed in a ``top-down''  process that prioritizes clinical expertise over user input \cite{bhattacharjee2023investigating, kabir2022ask}. While mental health experts possess knowledge of mental illness and evidence-based treatment strategies, they may not have a complete understanding of users' experiences, self-management approaches,  and habits regarding technology.} Consequently, expert-designed tools may not meet users' needs or be compatible with their daily routines \cite{kornfield2022meeting, bhattacharjee2022connecting}. 


To overcome the limitations of prior gratitude-promoting tools, we posit that a user-centered design (UCD) approach \cite{abras2004user, mao2005state, gulliksen1998user, kornfield2022meeting} can help identify the key features necessary to help young adults incorporate gratitude practice into their daily routine.  UCD is an iterative design process that seeks to create products that better align with users' needs and preferences. By engaging users in the design process and comprehending their needs, preferences, and goals, we sought to develop a tool that not only fosters gratitude but also resonates with their realities and experiences.  

Among the various platforms available to promote gratitude, such as web applications and virtual reality (VR) technologies, we opted to develop a mobile application. This choice was driven by the ubiquity and widespread usage of smartphones, particularly among young adults \cite{heron2019mobile, mobilestat}. Mobile applications offer a convenient and accessible means of engaging with users, as they can be easily installed on smartphones, which are commonly carried by individuals throughout the day. Developing a mobile application also provided us with the flexibility to incorporate customized features specifically designed to promote gratitude. We recognized that the unique capabilities of mobile applications, such as push notifications, personalized user interfaces, and interactive experiences, might enhance user engagement and facilitate the practice of gratitude more effectively than other platforms.

\revision{This study's primary goal was to develop a mobile application that fosters gratitude practice, specifically designed to align with the needs and preferences of young adults, utilizing a UCD approach. Our focus was not on crafting novel features distinct from other applications; rather, we aimed to develop an application that genuinely reflects users' needs and experiences in digital gratitude practices. We emphasized iterative development of the application through close collaboration with young adults, differing from traditional ``top-down'' approaches that often prioritize expert input over user feedback. } Our study was motivated by the following two research questions:

\begin{itemize}
    \item \textbf{RQ1:} What are the key features of a mobile application that could promote regular gratitude practice among young adults?
    \item \textbf{RQ2:} What are the perceived benefits of using such applications among young adults, and how do these benefits vary based on users' engagement levels?
\end{itemize}


Our investigation began with a formative study involving interviews with 20 young adults who used two highly-rated gratitude applications on their mobile phones for a period of two weeks. During this phase, participants shared insights on their individual preferences for structuring their expression of gratitude and the importance of prompts during the evening for reflection and mood labeling. This feedback informed the design of our subsequent mobile application, which featured two versions: one incorporating predefined categories for gratitude entries (experiment condition) and another without such categorization (control condition).

The application was then deployed to 26 young adults, and our analysis indicated that participants in the experiment condition may have had higher levels of engagement, as evidenced by their higher word counts in comparison to those in the control condition. The qualitative data collected through interviews reinforced the benefits of providing structured options and evening prompts, highlighting their positive impact on the users' gratitude practice. However, participants also expressed concerns about the frequency of mood labeling, finding it to be overwhelming at times.
The deployment study also demonstrated that participants experienced positive shifts in their perception of the world and extended their gratitude practice beyond the application, incorporating it into their daily lives. Even in periods of passive engagement with the application, participants derived benefits from simply contemplating the positive aspects of their lives, resembling the benefits associated with active engagement.

In summary, our contributions include:

\begin{itemize}
    \item Identification of key features for a gratitude-promoting mobile application among young adults.
    \item Observation of the benefits experienced by young adults through regular gratitude practice using the application.
    \item A set of design considerations for gratitude-promoting applications for young adults, such as balancing the frequency of mood labeling questions and allowance for passive engagement.
\end{itemize}

%% file: Sections/related_work.tex
\section{Related Work}


This section begins by highlighting the positive impact of practicing gratitude on wellbeing. It then provides a concise review of various interventions developed by researchers to promote gratitude. Lastly, the section discusses the application of UCD principles in the development of digital mental health interventions.

\subsection{Positive Impact of Gratitude on Wellbeing}


Gratitude is an emotional and social experience that occurs when individuals acknowledge the positive aspects of their lives, including the people, situations, and experiences that bring them happiness, contentment, and wellbeing \cite{rash2011gratitude, emmons2004psychology, emmons2011gratitude}. It involves expressing appreciation for these positive experiences, either internally or externally, and recognizing the sources of goodness in one's life \cite{roberts2004blessings, watkins2004gratitude, weiner1985attributional}. Gratitude is considered a positive psychological trait that can be cultivated through various practices and interventions \cite{lyubomirsky2008happiness}.


Practicing gratitude has been extensively researched in the field of psychology, with numerous studies demonstrating its positive effects on wellbeing \cite{emmons2011gratitude, lyubomirsky2008happiness, alqahtani2022usability}. Several studies have shown that individuals who regularly express gratitude experience lower levels of anxiety, depression, and loneliness \cite{breen2010gratitude, brown2003measuring}. Gratitude has been found to promote positive interpretations of negative experiences, enabling individuals to cope with stress and trauma and even have increased resilience in the face of adversity \cite{emmons2011gratitude, fredrickson2003good, kumar2022does, shabrina2020gratitude}. Furthermore, practicing gratitude can help individuals savor positive life experiences, allowing them to extract maximum satisfaction and enjoyment from their current circumstances \cite{lyubomirsky2008happiness}. 

\begin{revisionenv}
Prior research has also shown that gratitude can play a significant role in building and strengthening social bonds between individuals \cite{algoe2008beyond, froh2010being, lyubomirsky2005benefits, lampinen2013indebtedness}. 
Expressing gratitude towards specific people, even if not communicated directly, is associated with increased closeness and stronger relationships with those individuals \cite{algoe2008beyond}. \citet{froh2010being} argued that gratitude can initiate a positive feedback loop, whereby strong relationships provide individuals with reasons to be grateful, in turn, strengthening those same relationships. Moreover, grateful individuals are more likely to exhibit positive traits, such as optimism and kindness, which are attractive to others and can lead to the formation of new friendships \cite{lyubomirsky2005benefits}.
\end{revisionenv}



In conclusion, regular expression of gratitude has been found to positively impact various aspects of individual and collective wellbeing. These benefits motivated the design and development of numerous interventions to promote gratitude, which we briefly describe in the next subsection.

\subsection{Interventions to Promote Gratitude}

Gratitude interventions have been implemented both offline and online, with the aim of improving wellbeing and social relationships. Offline interventions have included keeping gratitude journals, writing gratitude letters, and engaging in acts of kindness \cite{cregg2021gratitude, lyubomirsky2008happiness, watkins2015grateful}. Research has shown that these interventions can increase life satisfaction and social support \cite{jackowska2016impact, kerr2015can} while decreasing negative emotions, stress, and symptoms of depression and anxiety \cite{proyer2014positive, lyubomirsky2008happiness}.


Online gratitude interventions have emerged in diverse formats, including mobile applications, web-based journals, and social media platforms. Mobile applications provide an array of features that facilitate gratitude practices, such as daily reflection prompts, personalized reminders for consistency, and multimedia options that enable users to convey gratitude through photos, videos, or audio recordings \cite{choy2019listen, kloos2022appreciating, ghandeharioun2016kind, tang2022co}.  For instance, \citet{kloos2022appreciating} developed a gratitude application called ZENN, which features evidence-based gratitude exercises and persuasive elements such as daily quotes and tailored feedback. Furthermore, some applications include social components to foster a sense of community and connection among users, enabling them to share and engage in gratitude practices together \cite{alqahtani2022usability, mf2022ty}. Other applications that aim to support users' general wellbeing or manage negative emotions have also included gratitude journaling as a key feature, as observed in \cite{legaspi2022user, wiese2020design, helgert2022you, yamamoto2022digitalizing}.

Other digital platforms, such as social media and immersive technologies like augmented reality (AR) and virtual reality (VR), are increasingly being used to promote gratitude practice. Social media platforms, through dedicated gratitude groups and hashtags, provide users with opportunities to share their gratitude experiences with a wider audience, fostering a sense of community and belonging \cite{sciara2021gratitude, koay2020gratitude}. Similarly, AR and VR technologies hold great potential for creating immersive gratitude interventions, wherein users can engage in a more profound and realistic gratitude practice by interacting with virtual environments or objects \cite{bunn2022gogratitude}. However, it is important to note that while social media platforms and AR/VR technologies offer unique opportunities for promoting gratitude practice, they may not be as ubiquitous or accessible as mobile applications.

\revision{Although there has been a surge in the development of online tools to encourage gratitude practice, many of these tools still exhibit significant limitations in their design and functionality, not adequately addressing the varied needs and experiences of their end-users. Often, these tools are shaped primarily by academic and clinical perspectives (e.g., \cite{kloos2022appreciating, choy2019listen, ghandeharioun2016kind}), which may overlook the intricate preferences and real-world contexts of users. This oversight is critical, as many existing applications do not fully meet the diverse needs and contexts of individual users, leading to a lack of customization options, limited interactivity, or an absence of features that resonate with users' daily lives and personal practices \cite{kabir2022ask, muller2017conceptualization, kornfield2022meeting}. Acknowledging these limitations, recent literature on digital mental health (DMH) tools \cite{alqahtani2022usability, kabir2022ask} advocates for a more user-centered approach, engaging with end-users to understand their contexts and technology usage patterns. We adopt a user-centered approach in the design and development of a mobile application to support regular gratitude practice, aiming to address these gaps identified in existing literature.}


\subsection{User-Centered Design of Digital Mental Health Interventions}


UCD is an approach to design that prioritizes the needs, preferences, and behaviors of the end-users of a product or service \cite{abras2004user, mao2005state, gulliksen1998user, kornfield2022meeting}. In UCD, the design process is centered around the users, with the goal of creating products and services that are easy to use, effective, and satisfying for the people who will be using them. UCD involves gathering data on user needs, preferences, and behaviors through user research methods such as interviews, surveys, observations, and usability testing \cite{kornfield2022meeting}. This data is then used to inform the design of the product or service. Throughout the design process, UCD also emphasizes iteration and feedback. Designers create prototypes of the product or service and test them with users, gathering feedback on what works and what does not. This feedback is then used to refine the design and create a final product or service that meets the needs and expectations of the end-users \cite{kornfield2022meeting}.

UCD approaches have been increasingly utilized to develop interventions that meet the specific needs and preferences of target users, particularly those that are not adequately addressed by existing DMH tools \cite{liverpool2020engaging, kornfield2022meeting, kornfield2020energy, alqahtani2021co, kenny2016developing, o2017design}. For example, \citet{kornfield2022meeting} carried out design workshops to pinpoint essential features for automated text messaging tools in mental health interventions for young adults. They found that such tools should support users in exploring varied concepts and experiences while accommodating differing engagement levels based on individual preferences, availability, and mood, thus ensuring a tailored and adaptive user experience. 
Similarly, \citet{birrell2022development} surveyed and interviewed adolescents to develop a mobile application to provide peer support around anxiety, depression, and substance use.
\citet{stawarz2014don} analyzed user reviews of medication reminder applications and proposed design recommendations, such as habit formation and backup notifications. Overall, these studies highlight the importance of UCD approaches to developing effective digital interventions for young adults.

We observe that existing literature on user perspectives and needs regarding the promotion of gratitude among young adults is limited. While \citet{alqahtani2021co} identified several  strategies for general wellbeing applications (e.g., relaxation exercises for relieving symptoms of depression, distracting activities during periods of low mood) through a formative work and integrated some of these strategies into their subsequent gratitude application development \cite{alqahtani2022usability}, their focus was not exclusively on promoting gratitude. Our objective, however, is to concentrate solely on fostering gratitude; we seek to identify the critical features and design elements that resonate with user needs and perspectives,

%% file: Sections/overview.tex
\section{Research Overview}

\begin{revisionenv}

This research comprises two distinct but interconnected studies. The initial phase, referred to as the `Formative Study,' involved participants using an existing third-party gratitude application. The primary aim of this phase was to gather insights into the design features users anticipated would encourage regular gratitude practice in their daily routines. The insights gleaned from this formative phase shaped the next stage of our research.

\begin{figure}[h!]
    \centering
    \includegraphics[width=1\linewidth]{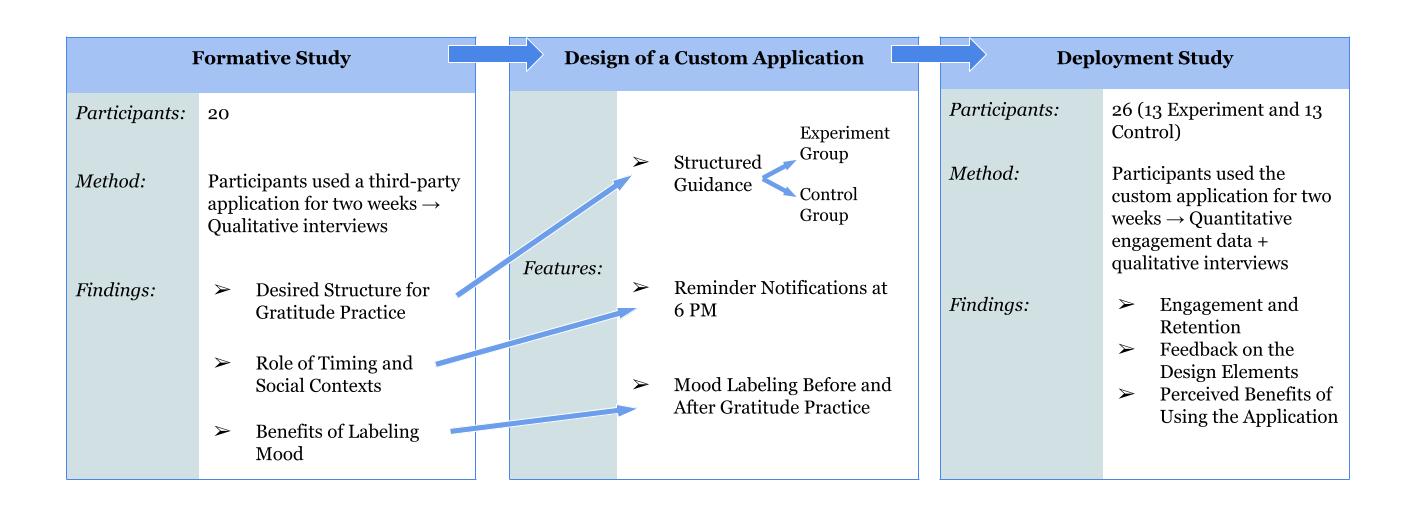}
    \caption{A High Level Overview of Our Work}
    \label{fig:diagram}
\end{figure}

The subsequent phase, the `Deployment Study,' involved the introduction and utilization of a custom-developed gratitude application. This study was designed to test and further explore the findings from the Formative Study. Participants interacted with our tailored application for two weeks and provided  feedback that confirmed, challenged, or expanded the initial findings. Figure \ref{fig:diagram} shows a high level overview of our work.
\end{revisionenv}

%% file: Sections/formative_procedure.tex
\section{Formative Study}
We conducted the formative study with 20 individuals to explore their anticipated design features that would facilitate regular gratitude practice in their daily lives. Below we describe how we set up the logistics for this study.

\subsection{Participants}
For this study, we recruited 20 participants from a large European and a large North American city through social media and email invitations. Since we aimed to promote gratitude practice among the general population, we did not impose any specific threshold criteria for mental health symptoms.
The inclusion criteria specified that participants had to be at least 18 years old and own a smartphone. The final sample consisted of 10 women and 10 men (other gender options were also offered), with a mean age of 21.2$\pm$0.78 years.
We refer to these participants as FP1--FP20.

\subsection{Procedure}


\begin{revisionenvsecond}
    
Our primary aim for conducting the formative study was to gain insight into the features and benefits that users would value in such an application. We understood that asking participants to comment on useful features and design ideas without any recent real-life experience of using gratitude applications could be a difficult task \cite{bhattacharjee2023investigating, liao2022connecting}. Therefore, we requested them to use a third-party application in a self-directed way for a two-week period to elicit more substantial and informed feedback. 
\end{revisionenvsecond}

    

\revisionsecond{Our research team reviewed around 50 applications available for download from the Apple App Store and the Google Play Store. Among these, several applications were discarded because their free versions offered limited functionality, or because the gratitude functionality was integrated into larger applications designed to support a range of wellbeing practices, which could detract from our study's specific focus. Additionally, we dismissed certain applications due to performance issues and instability identified during initial testing. Eventually, our research team selected \textit{Delightful - Three Good Things} for Android and \textit{Gratitude 365} for iOS, as they aligned with our selection criteria.} These criteria were strategically chosen based on best practices in application design and user experience \cite{aldayel2017challenges, albert2022measuring, bate2023bringing}, which included:

\begin{itemize}
    \item \textbf{Simplicity:} To facilitate ease of use and maintain focus on the core functionality of gratitude journaling.
    \item \textbf{Gender Neutrality:} To ensure inclusivity and appeal to a diverse user base, particularly important in studies involving diverse demographics.
    \item \textbf{High Ratings:} To leverage applications that are well-received and validated by users, indicating a positive and effective user experience.
    \item \textbf{Free Availability:} To guarantee accessibility for all participants, eliminating financial barriers and enhancing the diversity of the study sample.
    \item \textbf{Comparable Design:} \revisionsecond{To ensure a consistent user experience across applications on both iOS and Android platforms, minimizing variability in interaction. At the time of our study, we could not find a single application available on both platforms that also satisfied all the other criteria listed already.}
\end{itemize}



\begin{figure}
    \centering
    \includegraphics{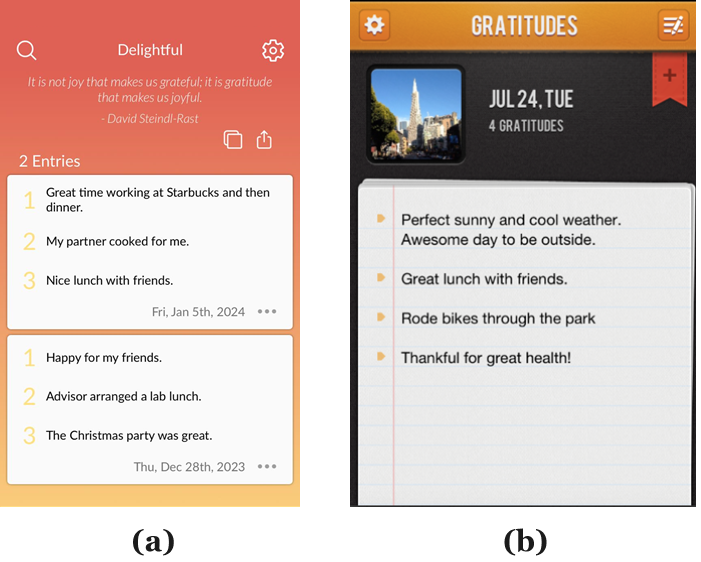}
    \caption{User interfaces of (a) Delightful -- Three Good Things and (b) Gratitude 365}
    \label{fig:apps}
\end{figure}

At the time of the study, both of the selected mobile applications allowed for the creation of multiple daily text entries at users' convenience. Although neither of them offered explicit guidance on the method or importance of practicing gratitude, \textit{Delightful -- Three Good Things} sent daily motivational quotes. A key feature of both applications was the ability to track and review past entries, displaying the total number of entries and their specific contents with date stamps.  Both applications included the feature of setting reminders to aid users in developing a routine of daily entries. The interfaces of both applications are presented in Figure \ref{fig:apps}. We did not offer any specific instructions on how to use these applications, nor were we able to track their usage data through the third-party applications’ databases. Instead, we encouraged participants to use the applications in a self-directed manner for two weeks, enabling them to provide us with feedback and suggestions based on their real-life experiences, rather than conjectures made in a vacuum.

Participants were interviewed via the Zoom videoconferencing platform after the two-week period. The interviews were conducted by one member of the research team and were semi-structured, allowing for flexibility to follow up on relevant insights and observations. The interviews lasted 15-35 minutes and focused on identifying the facilitators and barriers to using the applications. The interview questions covered a range of topics, including but not limited to:

\begin{itemize}
    \item What specific features of the application did you find most helpful or useful during the two-week period?
    \item What features would help you develop a consistent gratitude journaling habit?
    \item Were there any features of the application that you found confusing or difficult to use?
    \item Are there certain social contexts or situations where you prefer to use the application (e.g., alone, with others, in the morning, in the evening)?
\end{itemize}

All participants received a compensation of \$10 USD for their participation in the study. The research activities were conducted at two universities and were approved by the Research Ethics Boards of both institutions.


\subsection{Data Analysis}
\label{form:analysis}
After collecting the qualitative data from interviews, a thematic analysis approach \cite{cooper2012apa} was employed to analyze the data. Two team members acted as coders, who initially reviewed all transcripts to get familiar with the data. The coders then proceeded to assign segments of the data to distinct codes through an open-coding process \cite{khandkar2009open}. The two coders first individually developed a preliminary codebook before coming together to discuss and create a shared codebook. This process took place across several meetings, during which overlapping codes were identified, code definitions were refined, and codes not central to the research questions were excluded. Next, each coder applied the shared codebook to a subset of the data and then met again to refine the codebook based on the data. This iterative process was repeated until the coders reached a consensus on the final codebook, which was then applied to separate halves of the data.

\subsection{Ethical Considerations}
\label{subsec:formative_ethical}

Our team consisted of faculty members and graduate students with expertise in human-computer interaction, clinical psychology, and cognitive science. Given the nature of our research, which focused on promoting psychological wellbeing, we were mindful of the ethical considerations involved. Throughout our work, we took measures to address these concerns.

To ensure participant comfort and safety, we informed them at the outset of the interviews that they had the option to skip any questions or discontinue the conversation if they felt uncomfortable at any point. Our interviewers were trained in the Columbia-Suicide Risk Assessment protocol \cite{posner2008columbia} and prepared to respond appropriately if participants expressed thoughts of suicide or self-harm. They were equipped to provide safety planning or make referrals to crisis services as necessary. However, none of these risks emerged during the study.

%% file: Sections/formative_results.tex
\section{Formative Study Findings}
Our formative work identified several themes related to the desirable features participants envisioned in gratitude-promoting applications. In the following sections, we will discuss each of these themes and their implications for our subsequent deployment study.


\subsection{Desired Structure for Gratitude Practice}
\label{subsec:structure}

\revision{\textbf{Findings:} A common theme from our participant responses was the need for structured support in the form of examples and explanations within gratitude applications. This need was especially pronounced among participants who were engaging in gratitude practice for the first time. They often found the open-ended nature of the applications challenging. The uncertainty about what to include in their gratitude entries was articulated by FP4:} 
\revision{\italquote{I didn’t really know what to put into it. The first day it was like `I'm happy I finished work early' and `I'm happy the sun is shining' then the next day I didn't really know what else to put in, so I might have put in the same thing or something really similar. ... If I’d had a better idea of stuff to put into it, maybe I would've used it more, but I didn’t really know what to put into it.} }

\revision{This uncertainty often led to a diminished belief in the effectiveness of their gratitude practice. As FP18 described:}
\revision{\italquote{I find it quite cringe-y at the beginning and thought what I was putting in it was not right and stuff, I kept thinking oh I’m not doing it right, is this what I’m meant to be saying.}}

\revision{To address these challenges, participants indicated that additional support in the application could significantly enhance their experience. They believed that a structured approach with examples and explanations would offer a more effective framework for recalling and appreciating daily moments of gratitude.  FP16 commented:}
\revision{\italquote{Sometimes, I know I'm grateful, but I can't put it into words. A few examples could spark that thought process. ... Seeing sample entries or having occasional guidance would have assured me that I was on the right track and perhaps opened up new avenues of thought.}}

\revision{Participants with prior experience in paper-based gratitude journaling reported fewer challenges due to their existing understanding of the practice. However, they, too, recognized the potential benefits of structured guidance in digital platforms. They acknowledged that even with a foundational understanding, the addition of examples and prompts could enhance their practice by introducing fresh perspectives. For example, FP19 mentioned experiencing a tendency to repeat entries over time and expressed that a nudge or a fresh perspective from the app would be beneficial in diversifying their gratitude practice.}


\revision{These findings collectively underscore the importance of structured guidance in digital gratitude journaling apps.  Participants suggested that such features could not only aid in initiating and maintaining a gratitude practice but also enrich the experience by providing diversity and depth to the reflections.}
\\\newline
\noindent
\textbf{Implications for Our Deployment:} 
\revision{Previous research has highlighted the effectiveness of providing structured guidance with examples and instructions to individuals in mental health and behavior change applications, helping them to better understand and adopt a psychological principle in their life \cite{bhattacharjee2022kind, o2018suddenly}.  This insight emerged as particularly relevant in our formative study, where participants indicated that structured guidance could greatly enhance their experience with gratitude-promoting applications. Such guidance, which might include instructions on what aspects of life they might focus on (e.g., positive relationships) and relevant examples, can be beneficial for both novices and seasoned practitioners of gratitude exercises. The added support might offer a sense of direction, mitigating the initial overwhelm associated with starting a new habit and helping to establish a foundation for regular engagement with the tool \cite{bhattacharjee4057942exploring}. However, many existing applications, including those used in our formative study, opt for a more open-ended approach to support journaling. They predominantly offer an open-ended interface, where users are prompted to reflect on positive aspects of their lives without much directional guidance. This flexibility can be advantageous, allowing for a more personalized and self-directed journaling experience, particularly for those who find structured methods too restrictive \cite{furnham2011literature}. To delve deeper into the benefits and potential drawbacks of structured guidance compared to no guidance, our deployment study participants were divided into two groups:}

\begin{itemize}
    \item \textbf{Experiment Group:} Participants were provided with options to select life areas they were feeling grateful for, \revision{accompanied by explanations and examples}, before making daily gratitude entries in the application. 
    \item \textbf{Control Group:} Participants were asked to make daily gratitude entries in the application without any options to select life areas.
\end{itemize}

\revision{Our aim was first to understand the impact of structured guidance in facilitating gratitude practice. Consequently, we chose not to include advanced personalization features in our application, such as tailoring content based on users' previous experiences with gratitude practices or creating separate pathways for novices and experienced practitioners. This decision was made to prevent adding undue complexity to our exploratory work.}


\subsection{Role of Timing and Social Contexts}
\label{subsec:time}
\textbf{Findings:} Participants highlighted the importance of having suitable contexts that provided them with the necessary time and space to engage with the gratitude applications.  They reported using the gratitude applications most often when alone, affording time for reflection; ``\textit{You have to focus on it, and if around people, I would have rushed it.}'' (FP17). 
\revision{Some participants expressed that being around other people could hinder the state of mind required for practicing gratitude. FP14 remarked:} 

\begin{revisionenv}
\italquote{I don’t think that doing it while at a table, with people or something like that felt like the right time to do it. Because I’d
rather do it when I’m either like in bed or relaxing.}

As a result, participants tended to use the applications after regular working hours,  specifically in the evening before going to bed. This time frame provided them with the opportunity to engage in reflection without distractions, ensuring sufficient time and privacy for their gratitude practice. 
FP17 saw this as a form of ``me time'' that allowed for more focused reflection. Participants also indicated the need for a routine to positively influence the use of the application. FP13 suggested: 

\italquote{I've seen that putting it into my routine has kind of made it easier to keep it up, and by doing that, I’ve recognized the benefits are way stronger.}

\end{revisionenv}

Lastly, participants expressed a preference for privacy while using the gratitude journaling application, as it was considered a personal and introspective activity. To ensure this privacy, they suggested an incognito darker color scheme for the application interface so that the mobile screen is not easily visible to people around.
\\\newline
\noindent
\textbf{Implications for Our Deployment:}
Prior research works \cite{bhattacharjee2022design, bhattacharjee2023investigating, park2015manifestation} show that young adults generally prefer afternoons and evenings for reflecting on positive aspects of life since they generally have more free time then and are not in a rush to get to work or school like in the morning. Our formative study participants also shared similar opinions.
Therefore, our prototype application used in the deployment study sent notifications at 6 pm everyday. This time was chosen as it is a period when people are more likely to have some free time after working hours, but not too late that they may be too tired. 

\revision{The decision to send notifications at a fixed time, as opposed to randomizing them within an evening window, was grounded in the principles of habit formation and behavioral consistency. Research in behavioral psychology suggests that consistent cues, such as a notification at the same time every day, can significantly enhance habit formation \cite{eyal2014hooked, stawarz2014don}. Furthermore, \citet{lally2010habits}'s work on habit formation emphasizes the role of consistency in cue-response routines in establishing new habits. By aligning the notification timing with the natural downtime of our target demographic and reinforcing a consistent daily routine, we, therefore, aimed to embed the gratitude practice more effectively into participants' daily lives to support sustained engagement and habit formation.}


Additionally, participants emphasized the importance of privacy while using the application, in line with previous research \cite{eamons2003counting, o2017design}. We incorporated their feedback by using a dark blue background for the application interface.







\subsection{Benefits of Labeling Mood}
\label{subsec:mood}

\textbf{Findings:} The majority of the participants expressed that their emotional states, both before and after practicing gratitude, would influence their motivation to use the application. Most participants reported experiencing a positive shift in mood as a result of logging an entry, even when their initial mood was negative. FP15 said,

\italquote{It kind of shifted me if I was having a bad day. Made me kind of, well actually, I am going to focus here on the good things. ... There was a very conscious shift in my thinking on the bad days. I became more aware of trying to focus on the positives.}

Participants indicated that engaging in regular mood labeling activity and reviewing them at a later time would facilitate their ability to recognize negative thought patterns and triggers that influence their thoughts and emotions. This heightened awareness would, in turn, make them more receptive to the practice of gratitude as they actively sought out positive aspects of their lives to document and reflect on.  \revision{FP6 elaborated:}
\revision{\italquote{Reflecting on my emotions could be like connecting the dots between my thoughts, emotions, and what triggers them. ... This might make me more conscious of the need to focus on the positive.}}

Additionally, participants like P12 thought that seeing potential improvements in mood while focusing on positive experiences could also serve as reinforcement, further encouraging individuals to continue practicing gratitude.
\\\newline
\begin{revisionenv}
\noindent
\textbf{Implications for Our Deployment:} Informed by our formative study, we decided to integrate brief mood labeling questions before and after the gratitude activity in our application. Drawing on previous research that underscores the benefits of mood labeling, such as heightened awareness of emotions and thought patterns aiding in a better understanding of mental states \cite{torre2018putting, huang2015emotion, kou2020emotion}, we aimed to mirror this approach. Studies like \cite{lee2019benefits, nezlek2019within} have utilized numerical scales for participants to rate various components of emotions, stress, and life satisfaction, demonstrating the effectiveness of quantitative measures in capturing emotional changes.

Motivated by these insights, we aspired to design our mood labeling question in a way that quantitatively reflects mood changes. However, mindful of the risk of overwhelming users with too many questions \cite{bhattacharjee2023investigating}, we opted for a more streamlined approach. We included a single, straightforward question for participants to rate their mood on a scale from 1 to 5, with 1 representing a very low mood, 3 being neutral, and 5 indicating a very high mood, both before and after the gratitude activity. This design choice was driven by the desire to effectively capture mood changes without overburdening participants with extensive questioning \cite{bhattacharjee2023investigating}.

\end{revisionenv}

%% file: Sections/deployment_procedure.tex
\section{Deployment Study}

Although our formative study provided insights into the desired features of gratitude applications for regular practice, it is essential to complement these findings with deployment studies to understand how these expectations translate into real-world experiences \cite{rhodes2013big, sheeran2016intention}. In line with this recommendation, we conducted a deployment study to confirm, refute, and expand upon the findings from our formative work. In this section, we will first outline the design of the application and then provide details on the logistics of the deployment study.

\subsection{Design of the Application}



Our formative study findings informed the design of the gratitude-promoting application, \revision{which was developed to be compatible with both Android and iPhone devices.} This application was deployed in the subsequent phase of our study.


The application's design was primarily informed by the outcomes of our formative study. \revision{To understand the impact of structure and guidance for facilitating gratitude practice, we adopted a randomized approach in dividing participants into two distinct groups (experiment and control) (based on Section \ref{subsec:structure} findings).} The experiment group participants were offered an opportunity to select from specific aspects of life before making their daily gratitude entries, a feature we refer to as \textit{life area options}. These options included `Physical Wellbeing', `Peace \& Calm', Energizing Moments', `Engagement/Flow', `Connection', `Accomplishment', `Meaning/Fulfillment', and `Other'.  The selection of life area options in our application is based on the PERMA model from positive psychology literature \cite{seligman2011flourish, conner2018everyday, lyubomirsky2008happiness}. This model identifies key dimensions — Positive emotion, Engagement, Relationships, Meaning, and Accomplishment — that collectively contribute to personal and communal flourishing \cite{forgeard2011doing}. Studies have demonstrated strong correlations between these PERMA elements and various measures of wellbeing, including physical health, job satisfaction, and overall life satisfaction \cite{kern2014assessing}. To aid user understanding, the app contained a 'More Info' link to provide additional details and examples related to these life areas. For example, the life area `Connection' contained the following explanation:

\italquote{\textbf{Connection:} Moments of kindness, warmth, support, respect, connection with any other beings, including strangers, loved ones, and animals. E.g., ``warm smile from a stranger'', ``supportive collaborative meeting'', or ``lunch with friends''.}

After selecting an option, participants were provided with a textbox where they could write in detail about their feelings of gratitude. In contrast, the control group participants were instructed to make daily entries without selecting any specific life areas. However, \revision{participants from both conditions were required to rate their mood on a 1 to 5 scale before and after making an entry in the application (based on Section \ref{subsec:mood} findings). This mood rating system was integrated based on the understanding that mood labeling can enhance self-awareness, allowing individuals to recognize and understand their emotional states more clearly \cite{torre2018putting,  huang2015emotion, kou2020emotion}.} By tracking mood changes associated with the gratitude activity, we aimed to objectively assess the impact of the practice on individual moods. 

\begin{figure}
    \centering
    \includegraphics[width=0.6\linewidth]{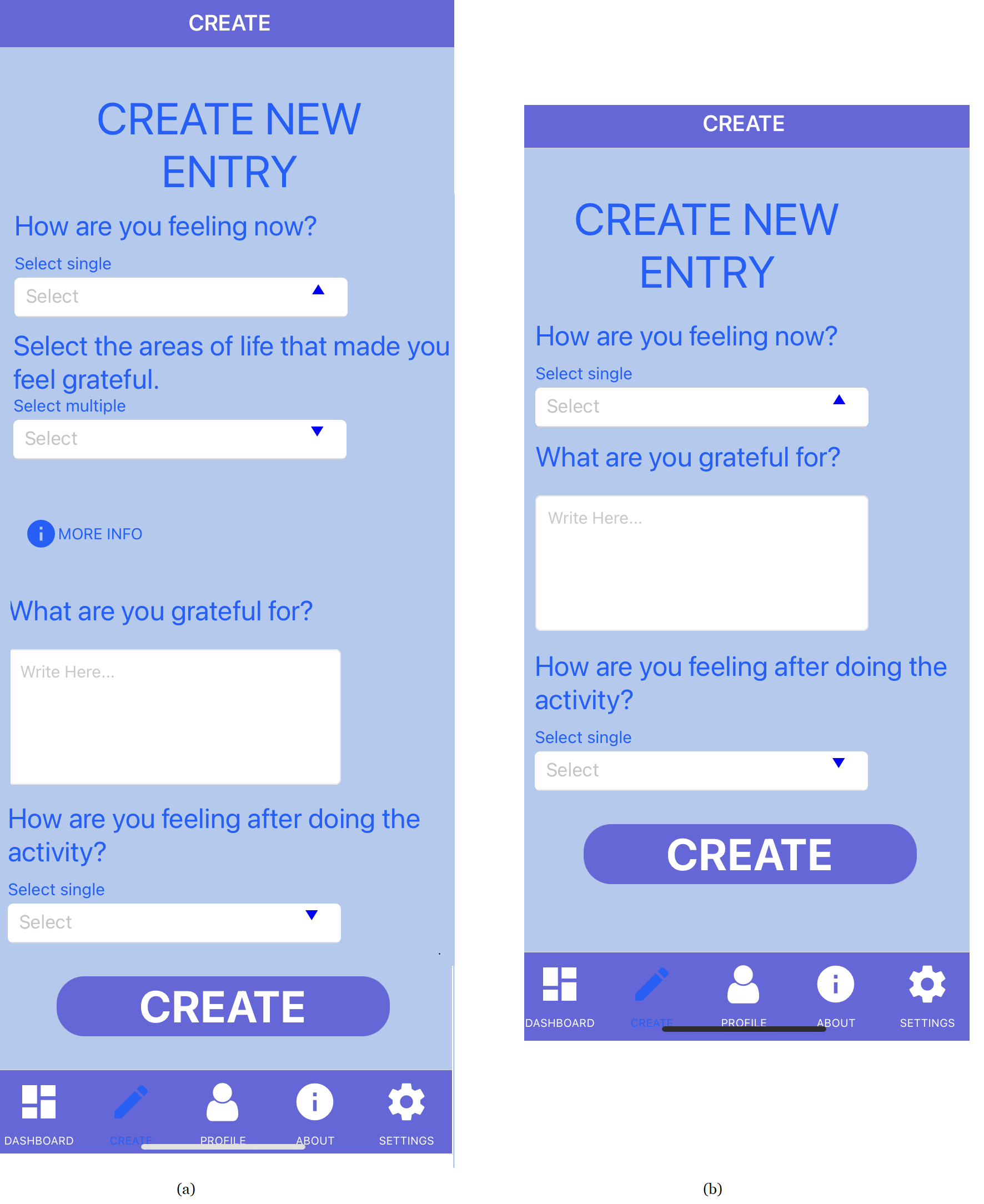}
    \caption{The interfaces for gratitude journaling in the (a) experiment  and the (b) control condition}
    \label{fig:exp_control}
\end{figure}

\begin{figure}
    \centering
    \includegraphics[width=1\linewidth]{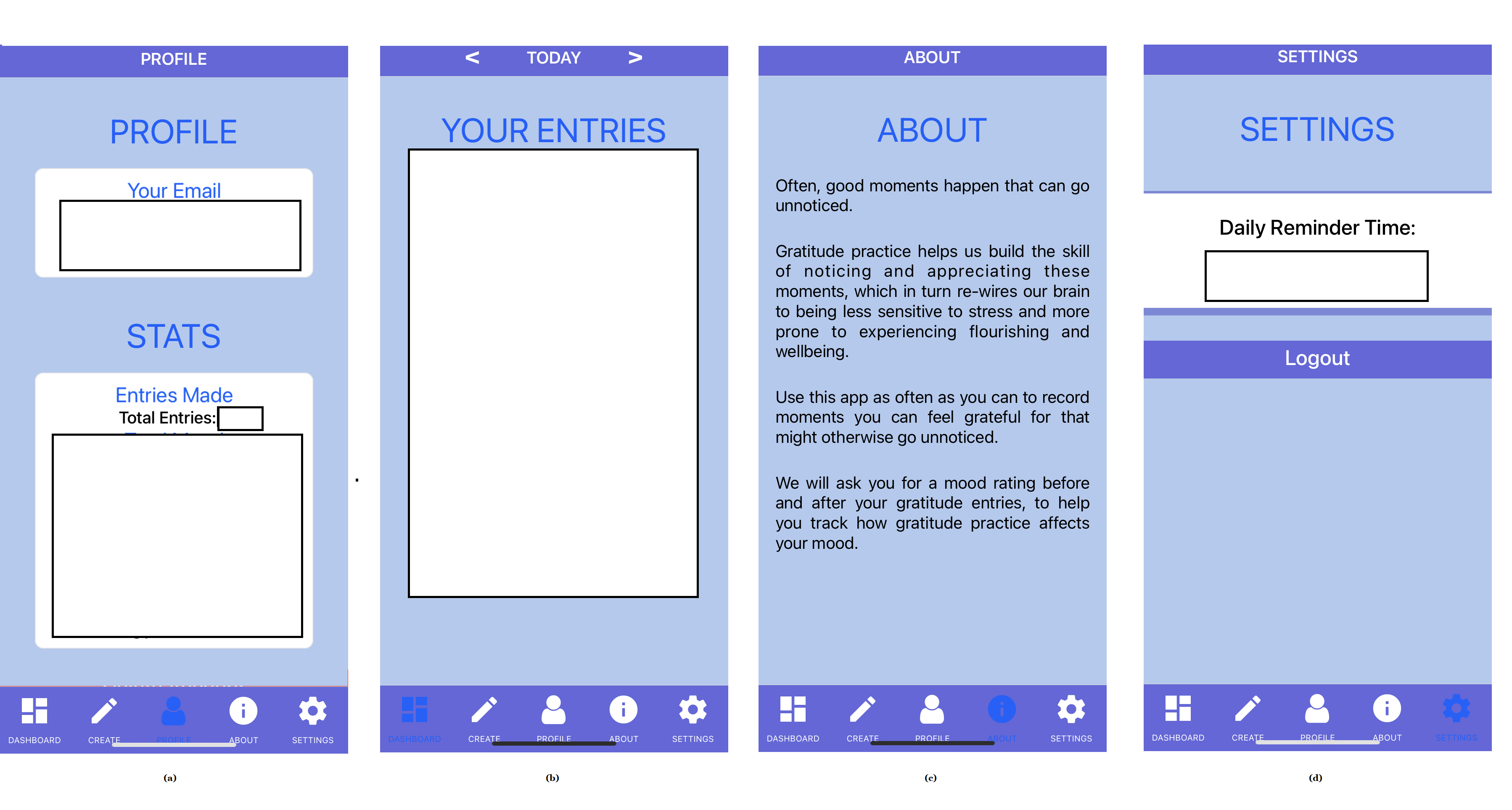}
    \caption{Screenshots of application interface showing (a) Dashboard page, (b) Profile page, (c) About page, and (d) Settings page. User-specific information has been obfuscated using a white rectangle for privacy.}
    \label{fig:interfaces}
\end{figure}

Figure \ref{fig:exp_control} illustrates the differences in the ``Create'' page interface within our application, highlighting how participants in both the experiment and control conditions interacted with the platform to record their gratitude experiences. We note that with the exception of the ``Create'' page, the application offered consistent functionality across both conditions. The ``Dashboard'' page enabled participants to access their previous entries, while the “Profile” page displayed their basic information and a comprehensive overview of their entries. The ``About'' page provided information on the benefits of
practicing gratitude and instructions on application usage, while the ``Settings'' page allowed participants to check their
daily reminder time and log out of their profile. Figure \ref{fig:interfaces} depicts the interfaces of various pages within the application.

\revision{The application sent daily reminders at 6 pm, prompting users to make an entry in the application (based on Section \ref{subsec:time} findings). Participants had the flexibility to make entries at any time of day, and the interfaces for making entries differed between the experiment and control conditions.}

\subsection{Participants}

Twenty-six participants were recruited for this study, with 13 randomly assigned to the experiment group and 13 to the control group. The inclusion criteria for recruiting participants remained consistent with the formative study, and participants were not provided information about the specific group assignments. We did not recruit any participants from the formative study in this phase.

Recruitment was conducted through social media and email invitations in a large European and a large North American city. The participants' mean age was $22.3\pm0.98$ years old, and they represented multiple genders (16 women, 9 men, and 1 non-binary/third gender). The control group is referred to as DP1c--DP13c, and the experiment group as DP1e--DP13e.

\subsection{Study Procedure}

Upon obtaining participants' consent to participate in the study, detailed instructions were provided to guide them through the installation process of the application on their smartphones, including both Android and iPhone devices. 
One member of the research team coordinated with the participants to ensure that the application was correctly installed on their phones. Then participants continued to use the application for the next two weeks. After the two-week period, participants were requested to fill up a concluding survey and subsequently invited to an interview session. In the concluding survey, 
we asked participants to provide their responses to the following questions using a 1-7 Likert scale, ranging from 1 (extremely low) to 4 (neutral) to 7 (extremely high), to assess their perceived motivation, engagement, and usefulness in practicing gratitude through the application:

\begin{itemize}
    \item How motivated did you feel to practice gratitude using the application? (Perceived motivation)
    \item How willing did you feel to engage in practicing gratitude when you were not in a good mood? (Perceived engagement)
    \item How useful did you find practicing gratitude using the application? (Perceived usefulness)
\end{itemize}


All participants were subsequently invited to take part in an interview session to provide their feedback on the application. Our interview questions included, but were not limited to:

\begin{itemize}
    \item How did you feel about being prompted to practice gratitude everyday?
    \item Did you notice any changes in your mood after you engaged in the gratitude activity? If so, when did the activity improve your mood, and when did it not?
    \item You may recall that we provided you with several options to choose from when selecting the topic for your gratitude entries. How did you feel about having those options available to you? (This question was only asked to participants in the experiment group.)
    \item What features might we add/remove to help you practice gratitude?
\end{itemize}

The interviews, which were done via the Zoom videoconferencing platform, took 20-45 minutes. All participants received a compensation of \$10 USD for their participation in the study.


\subsection{Data Analysis}


We computed descriptive statistics for the quantitative measures obtained from the concluding survey. Furthermore, we compared the average word counts of gratitude entries made by the participants in the application, \revision{which can reflect their effort and time} \cite{chen2021exploring, wang2016modeling, wang2023metrics}. 
The qualitative data from the interviews were analyzed in the same way as described in Section \ref{form:analysis}.

\subsection{Ethical Considerations}

At the beginning of our study, participants were informed that our gratitude application was not intended to serve as a crisis service. To ensure the wellbeing and safety of participants, we provided them with contact information for various crisis services, such as crisis text lines and suicide hotlines. While we did not actively solicit suicide-related information during the study, given the open-ended nature of the gratitude entries, we recognized the unlikely possibility that participants might disclose such thoughts or behaviors. To address this, we implemented procedures to monitor the entries made by participants on a daily basis. If any entry indicated a risk of self-harm or suicide, our team members were trained to promptly reach out to the participant and follow the Columbia-Suicide Risk Assessment protocol \cite{posner2008columbia}. These same considerations were also applied during the interviews, as described in Section \ref{subsec:formative_ethical}. Again, no such risks emerged, and no follow-up assessment was needed.

%% file: Sections/deployment_results.tex
\section{Deployment Study Findings}

In this section, we first provide a summary of the quantitative results pertaining to engagement and perceived benefits. We then explore participants' reactions to the design elements identified in the formative study, as well as the benefits they experienced while using both versions of the application.

\subsection{Engagement and Retention}

Table \ref{tab:quant} presents a comparison of quantitative measurements between the experiment and control conditions. Participants in the experiment condition showed a higher word count per entry in the application and experienced a greater increase in mood per entry compared to the control condition. The experiment condition showed slightly higher mean scores for perceived motivation and engagement, and the same mean score for perceived usefulness. 

\begin{table}[h!]
    \centering
    \caption{Comparison of quantitative measurements between the experiment and the control condition}
    \label{tab:quant}
    \begin{tabular}{|c|c|c|}\hline
        Measure & Experiment & Control\\\hline
        Word count per entry in the application & $15.27 \pm 2.42$ & $8.90 \pm 0.83$ \\\hline
        Mood increase per entry in the application & $0.64 \pm 0.04$ & $0.56 \pm 0.05$\\\hline
        Mean perceived motivation (1 to 7 scale) & $4.46 \pm 0.40$ & $4.38 \pm 0.40$\\\hline
        Mean perceived engagement (1 to 7 scale) &  $4.62 \pm 0.46$ & $4.08 \pm 0.44$\\\hline
        Mean perceived usefulness (1 to 7 scale) & $4.92 \pm 0.24$ & $4.92 \pm 0.35$\\\hline
    \end{tabular}

\end{table}

\begin{figure}[h!]
    \centering
    \includegraphics[width=0.8\linewidth]{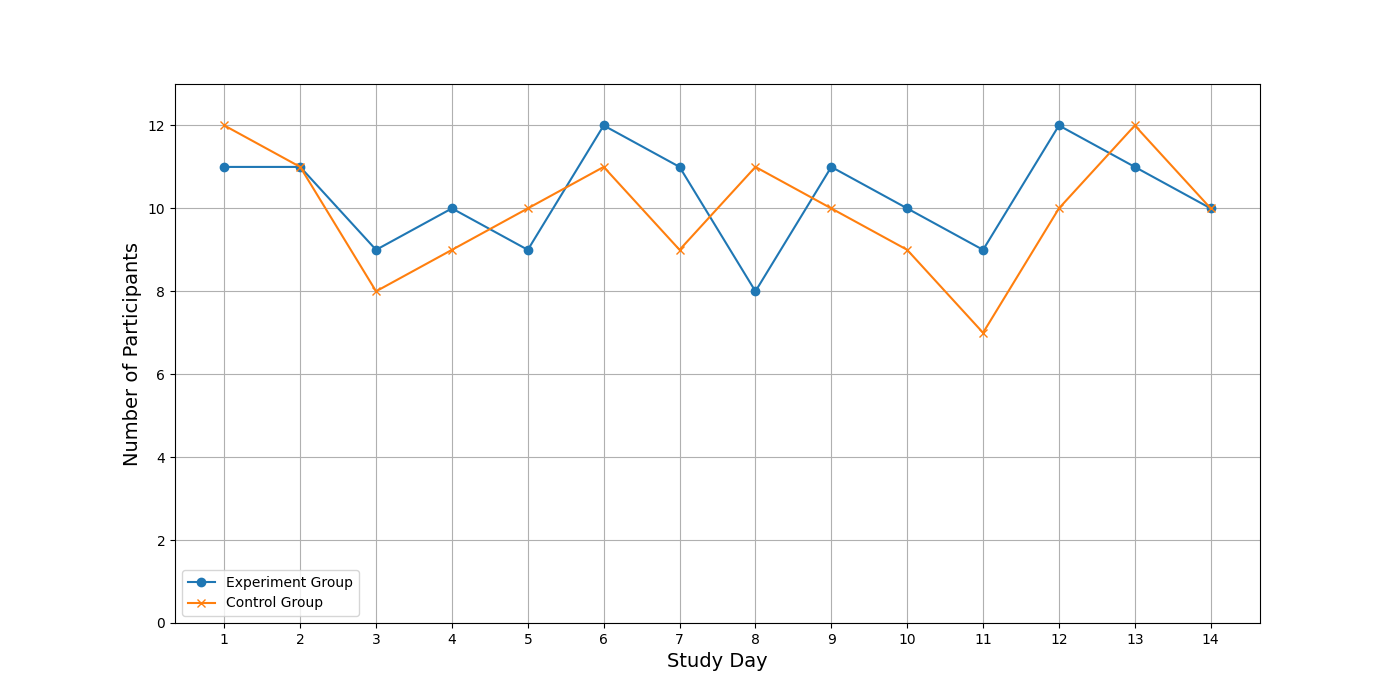}    
    \caption{Number of participants who logged entries in the application by study day for the experiment ($n=13$) and control ($n=13$) conditions}
    \label{fig:engagement}    
\end{figure}

\revision{Both conditions demonstrated comparable retention rates, with participants in the experimental condition logging entries on 79.1\% of the study days in total, compared to 76.4\% in the control condition. This comparability in the number of active participants was maintained throughout the study period, as shown in Figure \ref{fig:engagement}. As detailed in our supplementary document, analysis of word count over time and between weekdays and weekends revealed that the experiment condition consistently yielded higher word counts compared to the control group. It is important to note that these statistics do not account for passive engagement. As mentioned in Section \ref{subsub:passive_eng}, participants sometimes engaged with the application passively, without actively making an entry. Hence, we refrain from making definitive claims from these promising results; rather, we include them to supplement and bolster our qualitative findings.}


\subsection{Feedback on the Design Elements}

Below, we discuss how the deployment study participants responded to the design elements identified from the formative study.

\subsubsection{Presence and Absence of Life Area Options}
\label{subsubsec:life_area}


Participants in the experiment group reported that the application's prompts and structure guided them toward reflecting on different life areas and identifying positive experiences within those areas. For example, DP1-e shared: 
\begin{revisionenv}
    
\italquote{The app's daily prompts and structure made me reflect on the 'Connection' life area, which actually led me to recognize the support I've received from my family all my life. Using the app kind of made me realize that I often take my family's support for granted and didn't fully appreciate it. ... The daily habit of gratitude helped me develop a deeper appreciation for their love and support.}

\end{revisionenv}




This sentiment was echoed by several other participants, who  reported feeling more grateful and positive as a result of their participation in the study. Participant feedback also indicates that the application's life area options did not impose restrictions on their thoughts or gratitude entries. Instead, these predefined topics served as helpful cues that guided participants' reflections and focused their attention on specific aspects of their lives. According to them, the combination of structured prompts and open-endedness provided a balance that facilitated meaningful reflections while still allowing for personalization and individual expression. 
As DP13-e stated:

\italquote{I mean, I thought, for example, the topics were useful, because the topics help you focus on the boxes, like you don't think ``Oh, what was in my life?''. It's like a genre, like a box interaction. Like, you have 6 [[Actually 8]] boxes, and which of the boxes you want to put yourself in, which is nice, but then there is the open-ended text that helps you ``Okay, these are the boxes, what in my life is related?''. Then you think about your life, and there was the gratitude part. I think I felt like there was the part that was most meaningful to me. … Yeah, I'd say initially yes, initial cues are helpful to kind of narrow it down to areas of your life, and then you think about it as being like `well done today', which was really useful.}


A few participants in the control group, who did not know about the design of the application in the experiment group, expressed a desire for more structure in their gratitude journaling experience. They mentioned that it was challenging to come up with positive life areas to reflect on without any guidance or prompts. For instance, DP9-c specifically suggested having questions or options to guide their thinking, indicating that such structures would have been useful.

\subsubsection{Timing of Notification Prompts}

Participants from both conditions expressed appreciation for receiving notifications in the late afternoon.  Moreover, they valued the consistency of the timing, with several expressing anticipation for the daily notification at 6 pm.

However, some participants reported that they did not always make a gratitude entry immediately after receiving the notification. Instead, they noted that the timing of the notification served as a helpful reminder to make an entry when they had a free moment later in the evening. Many of these participants were engaged in family or dinner preparations when they received the notification, but the reminder helped them to prioritize gratitude practice when they had a moment to themselves. For instance, DP12-e described the experience by stating:

\italquote{I get the notification, I shut it down and I keep working for 10 more minutes, or 15 more minutes. Or I'm in the process like, ``Okay, now I need to cook. I need to get my stuff.'' So I’m like, “Okay, I'm turning this down, I'm doing it later”. }

DP3-c and DP13-e pointed out that it might not be feasible to have a single time that is suitable for practicing gratitude every day. DP3-c suggested that the application could focus on the user's current activity instead. They commented:

\begin{revisionenv}
\italquote{With the notifications, it should remind me to do this, during a time or routine in my day where it should be easy to do. For example, I actually often try to do it on the train.}
\end{revisionenv}

DP13-e suggested utilizing the Focus feature on iPhones, which allows users to customize specific activities or tasks and temporarily silence notifications unrelated to those activities. By aligning with the user's Focus settings, the gratitude application's notifications may have a higher probability of capturing their attention during periods of enhanced focus and productivity, according to them.


\subsubsection{Mood Labeling Activity}

The activity of labeling mood received mixed feedback from participants across both conditions. On the one hand, many appreciated the activity and found it helpful for recognizing their state of mind. Mirroring the findings from the formative study. they indicated that such activity could increase awareness of one's thoughts and emotions. Additionally, the activity allowed participants to tangibly observe how practicing gratitude improved their mood.  DP7-c explained the process: 

\italquote{It happened at two levels. One, when I was thinking about it. First I entered how I'm feeling and then I thought about what I'm going to enter now, or sometimes even before opening the app, it just came into my head that ``Oh, I'm thinking this, like I'm feeling this''. And now I'm thinking about gratitude, and now I'm going to make this entry. That itself was the first level of it, that's when I felt some change. And then when I was typing it, when I was articulating it, and I was trying to type it as well as possible, that's when the second level shift happened.  }

On the other hand, some participants found it overwhelming to label their mood every time they made an entry in the application. They felt that it added an extra step to the process and that it made the task of journaling feel `more like a chore' (DP6-e). \revision{In suggesting alternatives,   DP3-c recommended that the question on mood following the activity should not be asked immediately, but rather sometime later. They explained:}

\begin{revisionenv}

\italquote{Once I'm done, maybe you can also check to see how I am feeling an hour later or something like that. That might be interesting to see, just to see if there are any sustained effects after the gratitude exercise.}

On a different note, DP12-c recommended retaining the question following the gratitude activity and removing the former one.
Several others preferred the option to label their mood on occasion, rather than every time they made an entry in the application. These diverse views suggest the need for customizable options within the application to accommodate individual preferences for mood labeling.
\end{revisionenv}

\subsection{Perceived Benefits of Using the Application}

Below, we discuss the benefits participants experienced while using the application.

\subsubsection{Positive Shifts in Perception and Increased Resilience}

Several participants across both conditions reported that practicing gratitude helped them to view their lives in a more positive light and to appreciate what they had, rather than focusing on what they lacked. They described how expressing gratitude for even the smallest things in life, such as a warm cup of coffee or a kind word from a friend, helped them to reframe their outlook and cultivate a more optimistic mindset. DP8-c said,

\italquote{I realized at the end of it that I have so many things to be grateful for. You know, I'm blessed in many ways, and I just take it all for granted. That was a bit of a wake-up call. ... So, it made me more happier knowing that, you know, actually I can count my blessings.}

Participants like DP7-e also mentioned that practicing gratitude helped them to acknowledge the fragility of life and appreciate what they have.  They anticipated that they would become better equipped to cope with setbacks, as they had cultivated a positive perspective and an understanding that challenges are a part of life's journey. In their opinion, these realizations could potentially lead to an increased sense of resilience, allowing them to be more prepared to face adversity.

\subsubsection{Benefits of Passive Engagement}
\label{subsub:passive_eng}
A few participants across both conditions mentioned that they did not always make a gratitude entry in the application after receiving the notification. However, they still acknowledged that the notification helped them to mentally reflect on the positive elements of their lives. For some, this sort of passive engagement was due to low energy or busy schedules, but they still recognized the benefits of passive engagement. As DP2-c commented, 

\italquote{Sometimes it's not immediately obvious to me what the added benefit is of actually typing it out versus just thinking about it. Because the moment I think that I feel happy, I feel like I got what I want to get out of it, but that doesn't really leave a trace.}

In line with the above findings,  participants like DP8-c and DP10-e expressed their preference for passive engagement with gratitude practice rather than actively recording their gratitude entries every time. DP8-c highlighted that they were sometimes feeling uneasy about sharing personal and confidential information within the application, yet still recognized the benefits of practicing gratitude mentally. These findings emphasize the significance of the act of mental reflection itself, as it can sometimes complement the actual process of recording gratitude entries.

\subsubsection{Forming Habits for Sustaining Gratitude Practice}

Participants across both conditions reported that consistently writing about positive elements in life over the course of the study not only helped them practice gratitude within the application, but also influenced their behavior outside of the application. DP7-c found that making gratitude entries after receiving a scheduled notification increased their ability to practice gratitude, ultimately motivating them to continue the practice even without notifications. Another participant, DP1-e, revealed that they incorporated gratitude into their daily conversations with their spouse, saying:

\italquote{[[After using the application]] I usually sit down and have a heart-to-heart talk with my husband. Like, what are you grateful for? I'm like, What am I grateful for? I tell mine, and he says his. ... It's something we have. We do like to look back and think of all the good things in life.}

\revision{Similarly, some participants reflected on how the individual practice of gratitude, initially a personal endeavor, could naturally evolve into a communal activity, reinforcing social bonds. For instance, participants like DP5-e and DP12-c noticed that their personal habit of gratitude journaling began to influence their social interactions. They reported that what started as an individual exercise soon became a collaborative activity among peers. DP5-e recounted:}
\revision{\italquote{I started discussing the things I wrote in my gratitude journal with my friends during our hangouts. Soon, they too began sharing what they were grateful for. It's turned into this cool thing we do together, reminding each other of the good stuff in our lives.}}

By consistently practicing gratitude, participants became more familiar with the practice and developed a habit of focusing on the positive aspects of their lives. This habit ultimately helped them apply the activity to their daily lives, promoting a positive mindset and a greater sense of wellbeing.

%% file: Sections/discussion.tex
\section{Discussion}

\revision{In this study, we explored essential features for digital gratitude practices in a mobile application, as identified by young adults. Utilizing a custom application that prioritizes the expressed needs of target users, we investigated these features' real-world impact on their daily lives. Our research contributes to HCI and CSCW literature by demonstrating how applications, informed by UCD principles, can support gratitude practices in young adults' daily lives. This user-centered approach offers valuable insights for future research on promoting gratitude, providing information about users' needs and expectations and thereby guiding the development of effective and contextually relevant gratitude applications.}


In our discussion, we begin by summarizing the main findings of our study and highlighting their contribution to the existing literature. We then present design recommendations based on these findings, providing practical implications for the development of gratitude-promoting applications. Finally, we conclude by acknowledging the limitations of our work. 
\subsection{Key Insights}

\subsubsection{RQ1: Key Features of an Application for Promoting Gratitude}


In our study, we identified several key features of gratitude-promoting applications that could effectively promote regular gratitude practice among young adults. One of these features was providing participants with the option to select life areas they felt grateful for, which helped them structure their thoughts and reflections. This approach also allowed individuals to identify aspects of their lives that they may not have otherwise considered. The benefits of structured reflection activities have been noted in previous HCI and psychology literature, with studies such as \cite{o2018suddenly, bhattacharjee2022kind, aurora2023exploratory, o2017design} reporting that structured prompts can help people generate novel insights about their own life situations. 

Our findings also indicate that prompting young adults to practice gratitude during times when they are not overwhelmed by workplace or study-related activities, such as in the late afternoon or evening, could be effective in capturing their attention \cite{bhattacharjee2022design, bhattacharjee2023investigating, park2015manifestation}. \revision{This strategy mirrors the principles used in productivity-enhancing systems, where tasks are scheduled at times most conducive to focus and engagement \cite{kaur2020optimizing, poole2013hci}. Furthermore, the effectiveness of consistent notifications in establishing habitual behavior, as noted by our participants, resonates with strategies employed in just-in-time adaptive interventions (JITAIs) \cite{kabir2022ask, nahum2018just}. These interventions, often used in productivity and health applications, deliver prompts or actions at times when users are most likely to be receptive, such as during their `idle times' for activities like checking messages, social networking platforms, or news \cite{stawarz2014don, eyal2014hooked, poole2013hci, rennick2016health}. By adopting a similar approach in gratitude practices, where reminders or activities are timed to coincide with these idle moments, future works can enhance user engagement and foster more consistent practice.}


Lastly, our study emphasizes the significance of reflecting on one's mood before and after gratitude journaling, consistent with prior literature highlighting the benefits of mood labeling \cite{bhattacharjee2023investigating, torre2018putting, huang2015emotion, kou2020emotion}. However, our deployment study also highlighted participant concerns regarding the requirement to label their mood during each gratitude entry and the number of mood labeling questions they had to answer. Some participants found it overwhelming to label their mood for every gratitude entry, as they were asked to do so each time and respond to two mood labeling questions.  Hence, our results indicate the need for personalizing the frequency and repetition of questions associated with the mood labeling activity to ensure a positive user experience \cite{alqahtani2022usability, kornfield2022meeting}.

\subsubsection{RQ2: Perceived Benefits}

Participants in both the formative and deployment studies expressed appreciation for the benefits of regular gratitude practice. The act of expressing gratitude helped them to reframe their outlook on life, cultivating a more positive mindset and an appreciation for even the smallest things. This finding aligns with prior research on the positive effects of gratitude \cite{emmons2011gratitude, lyubomirsky2008happiness, alqahtani2022usability}. Moreover, participants also reported that practicing gratitude helped them to acknowledge the fragility of life, potentially increasing their sense of resilience to face adversity \cite{fredrickson2003good, kumar2022does}. \revisionsecond{These findings underscore how a gratitude application can aid young adults during key transitions, like relocating, beginning a new job, or navigating changes in relationships.  The application's structured prompts can highlight consistent elements of support, fostering resilience amidst change. In moments where young adults might find it challenging to stay positive, engaging in gratitude practices may also improve mood, mirroring the positive mood shifts reported by our study participants.}


\revision{Our findings also hinted at the community and collaborative aspects of gratitude practice. Regular engagement in gratitude exercise could not only foster individual wellbeing but also contribute to enhancing interpersonal relationships. Participants reported that sharing their gratitude reflections with others, such as family members or close friends, could lead to shared reflection on life and a more profound sense of connectedness. This communal aspect of gratitude practice, where participants extended their reflections beyond personal contemplation to include others in their circle, highlights the potential of gratitude interventions to foster a supportive and empathetic community environment \cite{lyubomirsky2008happiness, siangliulue2015toward, lasota2020become, froh2010being}. Furthermore, the sharing of gratitude experiences encouraged participants to engage in meaningful dialogues about positive aspects of life. This aspect aligns with previous studies indicating that gratitude can create an atmosphere of mutual appreciation and support \cite{lampinen2013indebtedness}.}


Our findings suggest that maintaining passive engagement with gratitude journaling applications can still provide benefits to individuals, even when they are unable to actively engage due to ongoing life issues. This is consistent with previous research on digital interventions and social media \cite{burke2010social, bhattacharjee2023investigating, verduyn2021impact}, which has emphasized the potential advantages of passive forms of engagement. In our study, we observed that notifications served as prompts for participants to reflect on the positive aspects of their lives, even when they did not actively make gratitude entries. Recognizing the importance of passive engagement during periods of low mood or busy schedules can help prevent attrition and encourage users to continue using digital mental health tools \cite{kornfield2020energy, bhattacharjee2023investigating}. These insights emphasize the need for flexible and personalized engagement strategies in the design of such tools to cater to the diverse needs and preferences of users.

\revision{Finally, our work highlights the considerable advantages of digital gratitude journaling over conventional pen-and-paper methods.  \cite{newman2021comparing, bhattacharjee2022design, alqahtani2022usability}. The application's interactive and personalized features, such as mood tracking and structured life area options, provided users with a more tailored user experience \cite{bhattacharjee2023investigating}. Scheduled notifications in the application acted as regular reminders for users to engage in gratitude practices \cite{bhattacharjee2022design}.  The application's capacity for immediate and convenient entry motivated regular practice too \cite{alqahtani2022usability}. These features, unique to a digital platform, could not only  make the practice of gratitude accessible and engaging but also amplify its positive impact on psychological wellbeing and interpersonal relationships.}


\subsection{Design Recommendations}
Our findings allowed us to make some design recommendations for the applications that aim to promote gratitude among young adults. We describe them below.

\subsubsection{Balancing the Frequency of Mood Labeling Questions}

Our deployment study participants provided diverse opinions regarding the two questions asked about their mood as part of the gratitude activity. While participants appreciated the opportunity to reflect on their mood and found value in this self-awareness, some of them expressed that answering both questions became overwhelming over time. To address this feedback, future systems can adopt various approaches to enhance user experience and engagement.

One approach is to offer personalization options, allowing users to customize the frequency of mood-related questions based on their preferences and needs \cite{alqahtani2022usability}. This flexibility will empower individuals to tailor the questioning process to their comfort level, avoiding the feeling of being burdened with excessive inquiries \cite{bhattacharjee2022design}. Another potential approach is to delay the timing of mood related questions. \revision{Rather than immediately following each gratitude entry, the system can schedule mood assessments at intervals that are more conducive to reflection \cite{kloos2022appreciating}}. This might allow users to engage in the practice of gratitude without the immediate pressure of assessing their mood, creating a more natural and unobtrusive experience that may align with the flow of their daily lives.

\subsubsection{\revision{Allowance for Varied Forms of Engagement}} 
Our findings suggest that it may not always be realistic to expect high levels of active engagement with DMH tools like ours. Therefore, future systems should explicitly provide opportunities for passive engagement to accommodate users' varying levels of availability and willingness to actively participate \cite{burke2010social, bhattacharjee2023investigating, verduyn2021impact}. One approach to facilitate passive engagement could be by communicating to users that during times of high stress or busy work, they can mentally engage with the practice of gratitude, without the need to actively write about it in the application. This can be supported by incorporating low-effort features, such as a button click or checkbox, to indicate that the user has practiced gratitude in their mind. By allowing for minimal effort, DMH tools can still capture valuable information about users' engagement while accommodating their contextual constraints.

We note that facilitating passive engagement does not mean discouraging active engagement. Instead, it is about finding a proper balance between active and passive engagement to provide users with the flexibility to extract the potential benefits of DMH tools \cite{bhattacharjee2023investigating}. \revision{To enhance active engagement, especially in a predominantly text-based application, incorporating multimedia elements such as images and videos could be beneficial \cite{choy2019listen, boggiss2020systematic}. This approach would acknowledge that sometimes, users may find it easier and more engaging to upload a visual representation of their experiences, rather than composing a detailed text entry. Additionally, incentivizing active participation could foster sustained engagement \cite{jardine2023between}. Techniques like providing periodic summaries of users' progress or introducing gamification elements, such as maintaining streaks for consistent entries, can motivate users to engage more regularly \cite{hamari2014does, kocielnik2018reflection}. For instance, acknowledging milestones like ``consecutive entries for X days/weeks'' can create a sense of achievement and encourage continued use.}

\revision{However, it is crucial to ensure that these incentives do not become counterproductive. Overemphasis on streaks or excessive notifications might lead to undue stress, counteracting the intended benefits of the tool \cite{etkin2016hidden, bekk2022all}. Therefore, the application should carefully balance encouragement with mindfulness of potential stressors. By offering a spectrum of engagement options — from passive to active and text to multimedia — future systems can empower users to choose how they interact with the tool based on their current needs and preferences.}

\subsubsection{Transferring Gratitude Practices to Real-Life Contexts}

Our findings indicate that engaging in regular gratitude practice through the application can serve as a motivator for individuals to express gratitude in other areas of their lives. For example, DP1-e reported that they extended the habit of writing about gratitude in the application to engaging in active dialogue with their spouse about things they were grateful for. This highlights the potential for applications to inspire users to incorporate gratitude into their real-life interactions and relationships \cite{alqahtani2022usability}, \revisionsecond{potentially aiding in the management of mental health challenges associated with life stage transitions}.
To further encourage users to practice gratitude in their daily lives, future systems could provide more concrete instructions and guidance. For instance, users could be prompted with narratives or stories illustrating how others facing similar challenges in life practice gratitude with their family members or peers \cite{bhattacharjee2022kind, burgess2019think, o2017design}. \revisionsecond{Additionally, guided questions could be offered to assist users in exploring ways to express gratitude and connect with supportive individuals in their network  \cite{o2018suddenly, meyerhoff2022system}}. However, it is important to acknowledge that such activities may require higher effort from users and hence be occasional rather than regular.

\revision{Recognizing the impact of gratitude practices on interpersonal relationships, the design of future gratitude applications should also emphasize their collaborative and communal potential. Integrating features that allow for social sharing, such as enabling users to share their gratitude entries with family and friends, can contribute to deepening connections and foster empathy \cite{ lampinen2013indebtedness, algoe2008beyond}. Additionally, incorporating group gratitude activities, like a shared digital gratitude board, could encourage communal participation and reflection \cite{wong2017giving}. Such applications could also include collaborative reflection prompts and storytelling features \cite{bhattacharjee2022kind, o2018suddenly, siangliulue2015toward}, enabling users to share personal experiences and engage in meaningful dialogues about gratitude. This communal approach, enhanced by guided discussion tools and notification reminders for shared activities, could not only strengthen individual wellbeing but also build a supportive and empathetic community environment.}


\subsubsection{Incorporating Individual Schedule and Activity Information}

Our decision to send prompts in the evening for gratitude practice was generally well-received by participants, as it often allowed them to engage with the application without the stress of rushing for work or school commitments. However, we recognize that the ideal time for gratitude practice can vary among individuals. To address this, future systems can provide users with the flexibility to set their own preferred time for prompts \cite{bhattacharjee2022design}. While this approach can cater to individual preferences, it is important to note that users may not always accurately predict their schedules or optimal engagement times.

\revision{To optimize the timing of gratitude prompts in interventions, integrating commonly used data sources such as digital calendars and mobile phone sensors, similar to JITAIs, can be highly effective  \cite{kabir2022ask, nahum2018just, kaur2020optimizing}. By analyzing digital calendar data to identify potential breaks in users' schedules, such as free time between meetings or classes, applications can pinpoint ideal moments for sending gratitude prompts \cite{howe2022design, eschler2015shared}. Gratitude applications can also utilize mobile phone sensors, such as accelerometer and Global Positioning System (GPS) sensor, which can provide contextual information about users' activities and whereabouts \cite{daskalova2020sleepbandits, rooksby2019student, farrahi2015predicting}. 
Collectively, these data sources not isolate users from their social contexts; rather, they should aim to understand and respect their social environments. 
For instance, the system could identify when a user is in a relaxed state or a conducive environment for reflection and then deliver gratitude prompts accordingly. Conversely, it might avoid times when the user is likely busy, such as during office hours or classes, inferred from their movement patterns and calendar data.
However, it is crucial to prioritize user privacy and implement strong measures to ensure that the monitoring of calendar and sensor data is done in a secure and privacy-conscious manner \cite{alqahtani2022usability}.}

\subsubsection{Designing for Users with Varied Levels of Gratitude Experience}

\revision{In our study, we found that both novice and seasoned practitioners of gratitude can benefit from guided instructions and explanations, yet we recognize a need for an additional layer of personalization to cater to users' varying levels of experience with gratitude practices. For individuals new to the practice of gratitude,  more hands-on instructions could be provided. This could take the form of an \textit{interactive gratitude journey} for novice users \cite{bhattacharjee2022kind}, guiding them through the different stages of gratitude practice. This journey would start with fundamental concepts and gradually introduce more complex elements, ensuring a smooth and supportive learning curve for newcomers. Such an approach would resonate with prior works that advocate for scaffolded learning experiences to build competence and confidence in new learners \cite{williams2016axis, slovak2016scaffolding, sobel2017edufeed}.}

\revision{For those with more experience in gratitude practices, the application could cater to their established habits by offering advanced reflection prompts and deeper analytical tools for tracking their gratitude journey \cite{kocielnik2018reflection, bhattacharjee4057942exploring}. To further enhance personalization, users could have the option to create custom prompts and set personalized practice goals, thus tailoring their experience to their specific needs and progress \cite{agapie2022longitudinal, baumer2015reflective}. These features have the potential to make the gratitude practice more dynamic and engaging for all users, regardless of their starting point.}


\subsection{Limitations and Future Works}

We acknowledge a couple of limitations in our study that warrant consideration. Firstly, while our application was designed specifically for young adults between the ages of 18-25, we conducted the study in a limited geographical scope, involving participants from only a European city and a North American city. This regional focus may restrict the generalizability of our findings to a more diverse population, as cultural and contextual factors can influence individuals' responses to gratitude practices \cite{de2017gender}.  \revision{Moving forward, future research should focus on exploring the cultural nuances in gratitude perceptions and develop interventions that align with the diverse values and cultural contexts of users.} 

\revision{Moreover, our study predominantly addressed the needs and expectations of young adults. It is important to recognize that other demographics, such as teenagers facing digital-age pressures \cite{orben2023social}, middle-aged adults navigating life transitions \cite{kobayashi2022acute}, and immigrants adjusting to new environments \cite{tachtler2020supporting}, also confront unique psychological wellbeing challenges. Their interaction with and expectations from gratitude practices could vary considerably, necessitating tailored approaches. Therefore, future research should broaden its scope to include these varied demographics, investigating their specific mental health needs and the effectiveness of gratitude practices in their contexts.}


Our study also relied primarily on self-reported measures, which are subject to certain limitations \cite{parry2021systematic, nevin2009low}. Self-reported measures may be influenced by response bias, social desirability, and participants' subjective interpretations. Future studies could incorporate additional objective measures, such as physiological data or behavioral observations, to complement the self-reported data and provide a more comprehensive understanding of participants' experiences.

\revision{Finally, although the features in our application in the deployment study were informed by insights from our formative study and relevant literature, we acknowledge that the design space we explored was not exhaustive. The application could have incorporated alternative approaches such as guidance through a simple chat application \cite{o2017design} or reflections on emotions through questions on stress and life satisfaction \cite{lee2019benefits}. There are also many personalization opportunities that could have altered the user experience, especially by adapting to the users' previous experiences with gratitude practices or allowing personalized reminders \cite{zhang2021designing}. However, given the exploratory nature of our study, we intentionally limited the application's complexity. Our primary objective was to understand the fundamental needs of users in digital gratitude practice. With our findings showing promise, we see these aspects of personalization as exciting avenues for future work.}


%% file: Sections/conclusion.tex
\section{Conclusion}

Regular practice of gratitude can help young adults manage their psychological wellbeing. In this work, we  employed a UCD approach to develop a mobile application that facilitates gratitude practice among young adults. Our formative study provided valuable insights into the preferences of users when it comes to expressing gratitude and the importance of prompts for reflection and mood labeling. Participants expressed a preference for engaging with the application after regular working hours, allowing for uninterrupted reflection. Building on these insights, our deployment study with the custom-designed application further validated the positive impact of structured life area options. Participants reported higher levels of engagement when they had the opportunity to focus their reflections on specific aspects of their lives, enabling them to identify and appreciate positive experiences within those areas. Additionally, the study revealed the benefits of passive engagement during busy periods, allowing participants to easily integrate gratitude practice into their daily routines without adding a significant burden. We look forward to our findings serving as a catalyst for future investigations regarding promoting gratitude among young adults.

\section*{Acknowledgements}

The authors acknowledge the use of ChatGPT, an AI language model developed by OpenAI, during the writing process of this paper. ChatGPT was employed to improve the text's readability in some parts of the paper. The authors retain responsibility for the content and interpretation of the findings. 